\documentclass[aps,twocolumn]{revtex4}%
\usepackage{amsfonts}
\usepackage{amsmath}
\usepackage{amssymb}
\usepackage{graphicx}%
\providecommand{\U}[1]{\protect\rule{.1in}{.1in}}
\begin{document}
\preprint{HEP/123-qed }
\title{Fisher information and spin squeezing in the Lipkin-Meshkov-Glick Model}
\author{Jian Ma}
\affiliation{Zhejiang Institute of Modern Physics, Department of Physics, Zhejiang
University, Hangzhou 310027 China}
\author{Xiaoguang Wang}
\email{xgwang@zimp.zju.edu.cn}
\affiliation{Zhejiang Institute of Modern Physics, Department of Physics, Zhejiang
University, Hangzhou 310027 China}
\date{\today }

\begin{abstract}
Fisher information, lies at the heart of parameter estimation theory, was
recently found to have a close relation with multipartite entanglement
(Pezz\'{e} and Smerzi, Phys. Rev. Lett. \textbf{102}, 100401). We use Fisher
information to distinguish and characterize behaviors of  ground state of the
Lipkin-Meskhov-Glick model, which displays a second-order quantum phase
transition between the broken and symmetric phases. We find that the parameter
sensitivity of the system attains the Heisenberg limit in the broken phase,
while it is just around the shot-noise limit in the symmetric phase. Based on
parameter estimation, Fisher information provides us an approach to the
quantum phase transition.

\end{abstract}

\maketitle

\section{Introduction}

Parameter estimation of probability distributions is one of the most basic
tasks in information theory, and has been generalized to quantum
regime~\cite{Helstrom,Holevo} since the description of quantum mechanics is
essentially probabilistic. How to improve the precision of parameter
estimation has been focused for many years, and is of important applications
in quantum technology like quantum frequency standards
\cite{Bollinger96,Huelga97}, measurement of gravity accelerations
\cite{Peters99}, and clock synchronization \cite{Jozsa00} etc.

Consider a quantum state $\rho_{\theta}=U_{\theta}\rho_{\text{in}}U_{\theta
}^{\dagger}$, where $U_{\theta}=\exp\left(  i\theta\hat{K}\right)  $ and
$\hat{K}$ is a generator. We estimate parameter $\theta$ through proper
measurements, however, the precision of our estimation is limited by the
quantum Cramer-Rao (QCR) bound \cite{Helstrom,Holevo},%
\begin{equation}
\Delta\hat{\theta}\geq\left(  \Delta\theta\right)  _{\text{QCR}}=\frac
{1}{\sqrt{\nu F\left(  \rho_{\text{in}},\hat{K}\right)  }},\label{QCR}%
\end{equation}
where $\nu$ is the number of trails, $\hat{\theta}$ is the so-called unbiased
estimator, and $F\left(  \rho_{\text{in}},\hat{K}\right)  $ is the quantum
Fisher information (QFI) \cite{Helstrom,Holevo,Hubner,Hayashi06}. In a sense,
parameter estimation is equivalent to distinguishing neighboring states along
the path in parameter space. We know QFI has close relation with Bures
distance \cite{Bures69}, the most studied distance in quantum-state space, and
Bures distance is directly related to the Uhlmann fidelity \cite{Uhlmann}. The
QFI is proportional to the Bures distance \cite{Wootters81,BraunsteinPRL72}.
For pure states, the QFI, as well as the Bures distance $ds_{B}^{2}$, is just
proportional to the variance of $\hat{K}$ \cite{Hayashi06}, that is $F\left(
\rho_{\text{in}},\hat{K}\right)  =4ds_{B}^{2}=4\left(  \Delta\hat{K}\right)
^{2}$. Therefore, besides increasing experimental times $\nu$, we can improve
the estimation precision $\Delta\hat{\theta}$ by choosing proper states
$\rho_{\text{in}}$ for a given $\hat{K}$. In general, entangled states are
more sensitive than separable states, i.e., the variance of $\hat{K}$ is
large. In the past, many works have been devoted to improvement of parameter
sensitivity by using entangled states
\cite{Yurke86,Dowling98,Kok04,Giovannetti04,Giovannetti06,Boixo07,Roy08,Boixo08,Hofmann09,Rosenkranz09,nature2008,grjin}%
.

Quite recently, Pezz\'{e} and Smerzi \cite{Pezze09} found an interesting
application of QFI in multipartite entanglement and the sub-shot-noise phase
sensitivity in the estimation of a collective rotation angle. Consider an
ensemble of spin-half particles in the state $\rho_{\text{in}}$, they
introduced a quantity%
\begin{equation}
\chi^{2}=\frac{N}{F\left(  \rho_{\text{in}},S_{\vec{n}}\right)  },
\end{equation}
and prove that $\chi^{2}<1$ implies multipartite entanglement. Here, the
generator of $\theta$ is $S_{\vec{n}}=\vec{S}\cdot\vec{n}$ that denotes the
collective spin operator along direction $\vec{n}$. Namely, a sufficient
condition is given for quantum entanglement. We may define a mean Fisher
information as $F_{m}=F\left(  \rho_{\text{in}},S_{\vec{n}}\right)  /N$. Then,
$\chi^{2}$ and $F_{m}$ are reciprocal to each other. The relation between
$\chi$ and QCR bound is%
\begin{equation}
\Delta\hat{\theta}\geq\frac{1}{\sqrt{F\left(  \rho_{\text{in}},S_{\vec{n}%
}\right)  }}=\frac{\chi}{\sqrt{N}}=\chi\left(  \Delta\theta\right)
_{\text{SN}},\label{CRB2}%
\end{equation}
where $\left(  \Delta\theta\right)  _{\text{SN}}=1/\sqrt{N}$ is the shot-noise
limit and we set $\nu=1$. Thus, it is evident that $\chi^{2}<1$ becomes a
necessary and sufficient condition for sub-shot-noise phase estimation.

In this work, we study the Fisher information of the ground state of the
Lipkin-Meshkov-Glick (LMG) model \cite{Lipkin65}, which has a second-order
quantum phase transition (QPT) \cite{sachdev}, between a symmetric (polarized,
$h\geq1$) phase and a broken (collective, $h<1$) phase. Some works have been
devoted to study the LMG model with concurrence \cite{DusuelPRB71} and entropy
\cite{VidalPRA69,Barthel06}. In our work we find that, besides indicating the
critical point and entanglement, $\chi^{2}$ reflects the performances of
ground states of these two phases in the sense of parameter sensitivity. In
the symmetric phase, $\chi^{2}$ approaches to 1 with the increasing of $h$,
and is independent of $N$, that means $\left(  \Delta\theta\right)  _{\text{QCR}%
}\sim\left(  \Delta\theta\right)  _{\text{SN}}$. In the broken phase, we find
$\chi^{2}\simeq1/N,$ thus $\left(  \Delta\theta\right)  _{\text{QCR}}%
\simeq1/N$ attaining the Heisenberg limit.

This paper is organized as follows. In Sec. II, we give brief discussions
about the relations between spin squeezing and $\chi^{2}$. Then in Sec III, we
study $\chi^{2}$ and spin squeezing for the ground state of the LMG\ model in
both isotropic $\left(  \gamma=1\right)  $ and anisotropic $\left(  \gamma
\neq1\right)  $ cases. In isotropic case, the LMG model is diagonal in Dicke
states. For Dicke states, we find that, $\chi^{2}$ and the spin squeezing
parameter by Kitagawa and Ueda \cite{kitagawa93} are reciprocal to each other.
In anisotropic case, we use Holstein-Primakoff transformation and derive
$\chi^{2}$ in the thermodynamic limit. The finite size behaviors of $\chi^{2}$
and the spin squeezing parameter in the critical point are also obtained. The
numerical results coincide well with the analytical ones.

\section{Fisher information and spin squeezing parameters}

Fisher information is related to spin squeezing, and there are two spin
squeezing parameters respectively given by Kitagawa and Ueda \cite{kitagawa93}%
, and Wineland \cite{Wineland94},
\begin{equation}
\xi_{1}^{2}=\frac{4\left(  \Delta S_{\vec{n}_{\perp}}\right)  ^{2}}{N},\text{
\ }\xi_{2}^{2}=\frac{N\left(  \Delta S_{\vec{n}_{\perp}}\right)  ^{2}%
}{|\langle S_{\vec{n}}\rangle|^{2}},
\end{equation}
where subscript $\vec{n}_{\perp}$ refers to an arbitrary axis perpendicular to
the mean spin $\langle\vec{S}\rangle$, where the minimum value of $\left(
\Delta S\right)  ^{2}$ is obtained. The inequality $\xi_{i}^{2}<1$ $(i=1,2)$
indicates that the state is spin squeezed. Spin squeezed states can be used to
reduce the measurement uncertainty \cite{kitagawa93,wineland92}, and improve
the measurement precision of the atomic clock transition
\cite{petrov07,Chaudhury06}. The spin squeezing inequality is a criteria for
multipartite entanglement \cite{Sorensen01,Wang03}. For an arbitrary
multiqubit separable states, it was found that $\xi_{2}^{2}\geq1,$ and thus
$\xi_{2}^{2}<1$ implies quantum entanglement.

As proved in \cite{Pezze09},
\begin{equation}
F\left(  \rho,S_{\vec{n}_{\bot}^{\prime}}\right)  \left(  \Delta S_{\vec
{n}_{\perp}}\right)  ^{2}\geq|\langle S_{\vec{n}}\rangle|^{2},\label{genunrel}%
\end{equation}
where directions $\vec{n}_{\bot}^{\prime},\vec{n}_{\bot},\vec{n}$ are
orthogonal to each other. $F\left(  \rho,S_{\vec{n}_{\bot}^{\prime}}\right)
=4\left(  \Delta\hat{R}\right)  ^{2}$, where $\hat{R}$ is determined by
$\hat{R}\rho+\rho\hat{R}=i\left(  \rho S_{\vec{n}_{\bot}^{\prime}}-S_{\vec
{n}_{\bot}^{\prime}}\rho\right)  $. In general, $\left(  \Delta\hat{R}\right)
^{2}\leq(\Delta S_{\vec{n}_{\bot}^{\prime}})^{2}$, and the equality is
obtained only for pure states. Then Eq. (\ref{genunrel}) reduces to the usual
uncertainty relation,
\begin{equation}
(\Delta S_{\vec{n}_{\bot}^{\prime}})^{2}\left(  \Delta S_{\vec{n}_{\perp}%
}\right)  ^{2}\geq\frac{|\langle S_{\vec{n}}\rangle|^{2}}{4},
\end{equation}
for pure states. The above inequality can be written in terms of the inverse
of the mean QFI and the squeezing parameter $\xi_{2}^{2}$ as%

\begin{equation}
\xi_{2}^{2}=\frac{N\left(  \Delta S_{\vec{n}_{\perp}}\right)  ^{2}}{|\langle
S_{\vec{n}}\rangle|^{2}}\geq\frac{N}{4(\Delta S_{\vec{n}_{\bot}^{\prime}}%
)^{2}}=\frac{1}{F_{m}}=\chi^{2}. \label{xi_chi}%
\end{equation}
Both the inequalities, $\xi_{2}^{2}<1$ and mean QFI $F_{m}>1$ ($\chi^{2}<1$),
indicate the presence of entanglement.

Furthermore, $\xi_{1}^{2}\leq\xi_{2}^{2}$, and there is no similar relation
between $\xi_{1}^{2}$ and $\chi^{2}$ like Eq. (\ref{xi_chi}). However, we find
that%
\begin{equation}
\xi_{1}^{2}\chi^{2}=\frac{\left(  \Delta S_{\vec{n}_{\perp}}\right)  ^{2}%
}{(\Delta S_{\vec{n}_{\bot}^{\prime}})^{2}}\leq1,
\end{equation}
since $(\Delta S_{\vec{n}_{\bot}^{\prime}})^{2}$ ($\left(  \Delta S_{\vec
{n}_{\perp}}\right)  ^{2}$) is the maximum (minimum) variance. As proved in
\cite{Wang03}, if the pure state is of exchange symmetry, $\xi_{1}^{2}<1$
implies entanglement. Then, from the above equation $\chi^{2}>1$ implies
entanglement. We know that $\chi^{2}<1$ indicates entanglement too. Therefore,
a pure symmetric state is entangled iff $\chi^{2}\neq1$ ($\xi_{1}^{2}\neq1$).

When the mean spin direction is along $z$ direction, the squeezing parameter
$\xi_{1}^{2}$ and $\chi^{2}$ becomes%

\begin{equation}
\xi_{1}^{2}=\frac{4\min\left\langle S_{\perp}^{2}\right\rangle }{N},~~\chi
^{2}=\frac{N}{4\max\left\langle S_{\perp}^{2}\right\rangle },
\end{equation}
where%
\begin{equation}
S_{\perp}=\cos\theta S_{x}+\sin\theta S_{y}.
\end{equation}
Furthermore, if $\left\langle \left\{  S_{x},S_{y}\right\}
\right\rangle =0$,
for instance, in the LMG\ model \cite{DusuelPRB71}, we have%
\begin{equation}
\xi_{1}^{2}=\frac{4\min\left(  \left\langle S_{x}^{2}\right\rangle
,\left\langle S_{y}^{2}\right\rangle \right)  }{N},~~\chi^{2}=\frac{N}%
{4\max\left(  \left\langle S_{x}^{2}\right\rangle ,\left\langle S_{y}%
^{2}\right\rangle \right)  },\label{sqz_fisher}%
\end{equation}
thus we only need to compute $\left\langle S_{x}^{2}\right\rangle $ and
$\left\langle S_{y}^{2}\right\rangle $ to determined the squeezing parameter
and quantity $\chi^{2}$ in the following discussions of QPTs in LMG model.

\section{Fisher information and squeezing in the LMG Model}

The LMG model, originally introduced in nuclear physics and has found
applications in a broad range of other topics: statistical mechanics of
quantum spin system \cite{BotetPRL49}, Bose-Einstein condensates \cite{Cirac},
or magnetic molecules such as Mn$_{12}$ acetate \cite{Garanin}. Recently, some
quantum-information concepts, such as quantum entanglement
\cite{VidalPRA69,Barthel06} and quantum fidelity \cite{SJGu08,Ma08}, have been
studied in this model, aiming at characterizing its QPT. It is an exactly
solvable \cite{PanPLB451,LinksPRA36}\ many-body interacting quantum system as
well as one of the simplest to show a quantum transition in the regime of
strong coupling. The quantum phase transition of this model is also clear and
interest: the ground state becomes degenerate and a macroscopic change in the
ground state energy takes place.

\subsection{LMG\ Hamiltonian}

The Hamiltonian of the LMG model reads%
\begin{equation}
H=-\frac{1}{N}\left(  S_{x}^{2}+\gamma S_{y}^{2}\right)  -hS_{z}, \label{lmg1}%
\end{equation}
where $S_{\alpha}=\sum_{i=1}^{N}\sigma_{\alpha}^{i}/2$ are the total spin
operators in the direction $\alpha=x,y,z$; $\sigma_{\alpha}^{i}$ are the Pauli
matrices; $N$ is the total spin number, $\gamma$ is the anisotropic parameter
and $h$ is the effective strength of the external field. Without loss of
generality, we assume $0\leq\gamma\leq1$ and $h\geq0$.

This system undergoes a second-order QPT at $h=1$, between a symmetric
($h\geq1$) phase and a broken ($h<1$) phase, which are associated with
single-particle and collective behaviors, respectively. These two phases are
well described by a mean-field approach \cite{DusuelPRB71}. The classical
state is fully polarized in the field direction $\left(  \left\langle
\sigma_{z}^{i}\right\rangle =1\right)  $ for $h\geq1$, and is twofold
degenerate with $\left\langle \sigma_{z}^{i}\right\rangle =h$ for $h<1$.
However, since the Hamiltonian is of spin-flip symmetry, i.e., $\left[
H,\prod_{i=1}^{N}\sigma_{z}^{i}\right]  =0$, we have
\begin{equation}
\left\langle S_{x}\right\rangle =\left\langle S_{y}\right\rangle
=0,~~\left\langle S_{x}S_{z}\right\rangle =\left\langle S_{y}S_{z}%
\right\rangle =0.
\end{equation}
Thus the mean spin direction is along the $z$-axis for finite size case. In
addition, $\left[  H,\mathbf{S}^{2}\right]  =0$, and the ground state lies in
the $S=N/2$ symmetric section.

\subsection{Isotropic case and Dicke state}

We begin with the simple isotropic case, $\gamma=1$. The Hamiltonian reduces
to%
\begin{equation}
H=-\frac{1}{N}\left(  \mathbf{S}^{2}-S_{z}^{2}\right)  -hS_{z},
\end{equation}
which is diagonal in the standard eigenbasis $\left\{  |S,M\rangle\right\}  $
of $\mathbf{S}^{2}$ and $S_{z}$. For $S=N/2$ the energy eigenvalue is
\begin{equation}
E\left(  M,h\right)  =\frac{2}{N}\left(  M-\frac{hN}{2}\right)  ^{2}-\frac
{N}{2}\left(  1+h^{2}\right)  ,
\end{equation}
and the ground state $|S,M_{0}\rangle$ is readily obtained when \cite{Ma08}
\begin{equation}
M_{0}=\left\{
\begin{aligned} &N/2 &\text{for}\quad&h\ge 1,\\ &N/2-R\left[N(1-h)/2\right] &\text{for}\quad&0\le h<1, \end{aligned}\right.
\label{eq:2}%
\end{equation}
where $R(x)\equiv\text{round}(x)$ gives the nearest integer of $x$. Then one
can see level crossings exist at $h=h_{j}$, where $h_{j}=1-\left(
2j+1\right)  /N$, between the two states $|S,S-j\rangle$ and $|S,S-j-1\rangle$.

As the ground state is actually a Dicke state $|S,M\rangle$, $\left\langle
S_{x}^{2}\right\rangle =\left\langle S_{y}^{2}\right\rangle =\left(
S^{2}+S-M^{2}\right)  /2$, then
\begin{align}
\chi^{2}  &  =\frac{N}{2\left(  S^{2}+S-M^{2}\right)  }\nonumber\\
&  =\frac{1}{N/2+1-M^{2}/N}\nonumber\\
&  \leq1, \label{iso_chi}%
\end{align}
the equality is obtained for $M=\pm S$. Immediately, we have%
\[
\xi_{1}^{2}=1/\chi^{2}\geq1.
\]
As we know that, when $M\neq\pm S$, the Dicke states are entangled but not
spin squeezed, since $\xi_{2}^{2}>\xi_{1}^{2}>1$. Numerical results of
$\xi_{1}^{2}$ and $\chi^{2}$ for the isotropic LMG model are shown in Fig.
\ref{fig1} (d). We can see that, in the broken phase, $M<S$, $\xi_{1}%
^{2}=1/\chi^{2}$, while in the symmetric phase, the ground state is
$|S,S\rangle$, thus $\chi^{2}=\xi_{1}^{2}=1$.

By considering $\chi^{2}$ in Eq. (\ref{iso_chi}), when $M$ is close to $\pm
S$, $\chi^{2}$ is just a bit lower than $1$, thus $\Delta\theta$ is not
improved much than $\left(  \Delta\theta\right)  _{\text{SN}}$. When
$\left\vert M\right\vert \leq\sqrt{S}$, we have
\begin{equation}
2/N\leq\chi^{2}\leq2/\left(  N+1\right)  ,
\end{equation}
and thus
\[
\left(  \Delta\theta\right)  _{\text{QCR}}=\chi/\sqrt{N}\sim1/N,
\]
which attains the Heisenberg limit. Although $|S,\pm S\rangle$ is not
entangled, "cat state" (or GHZ state)
\begin{equation}
|\psi\rangle=\left(  |S,S\rangle+|S,-S\rangle\right)  /\sqrt{2}%
\end{equation}
is entangled and useful in phase estimation \cite{Giovannetti04}. Under
$|\psi\rangle$, $\left\langle S_{\alpha}\right\rangle =0$, for $\alpha=x,y,z$,
thus there is no spin squeezing. We find the maximum variance $\left(  \Delta
S_{z}\right)  ^{2}=S^{2}$, then
\begin{equation}
\chi_{|\psi\rangle}^{2}=1/N,~~\left(  \Delta\theta\right)  _{\text{QCR}}=1/N,
\end{equation}
beating the Heisenberg limit. From the above analysis we know that, for
typical symmetry multipartite states, Dicke states, there are no spin
squeezing, while $\chi^{2}<1$ indicates that they are entangled and are useful
resources for phase estimation.

\subsection{Anisotropic case}

Now we consider the anisotropic case, $0\leq\gamma<1$. The spin expectation
values $\left\langle S_{\alpha}^{2}\right\rangle $ can not be obtained
analytically. By treating the quantum effect as small fluctuations,
approximate results can be obtained by using the Holstein-Primakoff (H-P)
transformation \cite{HolsteinPR58} in the thermodynamic limit, and by using
the continues unitary transformation method \cite{wegner,glazek93,glazek94}
for finite size case.

In the thermodynamic limit, the quantum fluctuations are small, we can use the
H-P approximation. This method requires one to determine the semiclassical
magnetization $\langle\vec{S}\rangle$, which is not along $z$-axis in the
broken phase in the thermodynamic limit. Following conventional steps, we
first employ a mean field approach, define a spin coherent state%
\begin{equation}
|\theta,\phi\rangle=%
{\textstyle\bigotimes\limits_{l=1}^{N}}
\left[  e^{-i\phi/2}\cos\frac{\theta}{2}|0\rangle_{l}+e^{i\phi/2}\sin
\frac{\theta}{2}|1\rangle_{l}\right]  ,
\end{equation}
under which%
\begin{equation}
\langle\theta,\phi|\vec{S}|\theta,\phi\rangle=\frac{N}{2}\left(  \sin
\theta\cos\phi,\sin\theta\sin\phi,\cos\theta\right)  .
\end{equation}
The Hamiltonian is rewritten as%
\begin{equation}
H=-N\left[  \frac{1}{2}\sin^{2}\theta\left(  \cos^{2}\phi+\gamma\sin^{2}%
\phi\right)  +h\cos\theta\right]  .
\end{equation}
As $\max\left[  \cos^{2}\phi+\gamma\sin^{2}\phi\right]  =\max\left(
1,\gamma\right)  =1$, ($\gamma\leq1$), we have%
\begin{equation}
\min H=-N\max\left[  \frac{1}{2}\sin^{2}\theta+h\cos\theta\right]  ,
\end{equation}
then we conclude: (i), symmetric phase, $h\geq1$, $\theta_{0}=0$, for all
$\gamma$; (ii), broken phase, $0\leq h<1$, $\theta_{0}=\arccos h$, $\phi
=0,\pi$, for $\gamma\neq1$. We emphasize that, the mean spin direction is
along the $z$-axis when the system is finite.

We rotate the $z$-axis to the semiclassical magnetization,%
\begin{equation}%
\begin{pmatrix}
S_{x}\\
S_{y}\\
S_{z}%
\end{pmatrix}
=%
\begin{pmatrix}
\cos\theta_{0} & 0 & \sin\theta_{0}\\
0 & 1 & 0\\
-\sin\theta_{0} & 0 & \cos\theta_{0}%
\end{pmatrix}%
\begin{pmatrix}
\tilde{S}_{x}\\
\tilde{S}_{y}\\
\tilde{S}_{z}%
\end{pmatrix}
.\label{rotate}%
\end{equation}
As presented in \cite{DusuelPRB71}, $\theta_{0}=0$ for $h>1$ so that
$\mathbf{S}=\mathbf{\tilde{S}}$, and $\theta_{0}=\arccos h$ for $h\leq1$. The
transformed Hamiltonian reads%
\begin{align}
\tilde{H}= &  -hm\tilde{S}_{z}-\frac{1}{N}\left[  \frac{m^{2}+\gamma}%
{2}\mathbf{\tilde{S}}^{2}-\frac{3m^{2}+\gamma-2}{2}\tilde{S}_{z}^{2}\right]
\nonumber\\
&  +\frac{h\sqrt{1-m^{2}}}{2}\left(  \tilde{S}_{+}+\tilde{S}_{-}\right)
-\frac{m^{2}-\gamma}{4N}\left(  \tilde{S}_{+}^{2}+\tilde{S}_{-}^{2}\right)
\nonumber\\
&  -\frac{m\sqrt{1-m^{2}}}{2N}\left(  \tilde{S}_{+}\tilde{S}_{z}+\tilde{S}%
_{z}\tilde{S}_{+}+\tilde{S}_{-}\tilde{S}_{z}+\tilde{S}_{z}\tilde{S}%
_{-}\right)  ,
\end{align}
where $m=\cos\theta_{0}$. Then we introduce the H-P transformation%
\begin{equation}
\tilde{S}_{z}=N/2-a^{\dagger}a,\text{~~}\tilde{S}_{-}=\sqrt{N}a^{\dagger}%
\sqrt{1-a^{\dagger}a/N}=\left(  \tilde{S}_{+}\right)  ^{\dagger}.
\end{equation}
The Hamiltonian can be written as%
\begin{align}
\tilde{H}^{\left(  0\right)  }= &  \frac{2hm-3m^{2}-\gamma+2}{2}a^{\dagger
}a-\frac{m^{2}-\gamma}{4}\left(  a^{\dagger2}+a^{2}\right)  \nonumber\\
&  +\frac{1-m^{2}}{4},
\end{align}
up to the $0$-th order of $N$. We neglect the terms of the $1$-th order of $N$
as they are constant. Now we use the Bogoliubov transformation%
\begin{equation}
a^{\dagger}=\cosh\left(  \frac{\theta}{2}\right)  b^{\dagger}+\sinh\left(
\frac{\theta}{2}\right)  b.
\end{equation}
To diagonalize $\tilde{H}^{\left(  0\right)  }$, we find%
\begin{equation}
\tanh\theta=\varepsilon=\frac{m^{2}-\gamma}{2hm-3m^{2}-\gamma+2}.
\end{equation}
%

\begin{figure}
[ptb]
\begin{center}
\includegraphics[
height=3.0388in,
width=3.775in
]%
{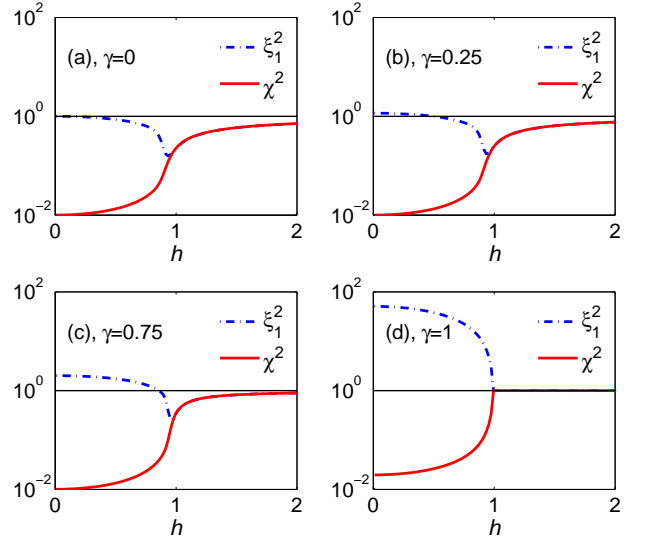}%
\caption{$\xi_{1}^{2}$ and $\chi^{2}$ as functions of $h$ for
various $\gamma $, with system size $N=100$. The crossing points of
$\xi_{1}^{2}$ and the horizontal line in the broken phase are
$h=\sqrt{\gamma}$.}%
\label{fig1}%
\end{center}
\end{figure}

The rotated spins are written under the H-P representation,%
\begin{align}
\tilde{S}_{x}= &  \frac{\sqrt{N}}{2}\left(  \frac{1+\varepsilon}%
{1-\varepsilon}\right)  ^{1/4}\left(  b^{\dagger}+b\right)  +O\left(
1/N\right)  ,\nonumber\\
\tilde{S}_{y}= &  \frac{i\sqrt{N}}{2}\left(  \frac{1-\varepsilon
}{1+\varepsilon}\right)  ^{1/4}\left(  b^{\dagger}+b\right)  +O\left(
1/N\right)  ,\nonumber\\
\tilde{S}_{z}= &  \frac{N}{2}+\frac{1}{2}\left(  1-\frac{1}{\sqrt
{1-\varepsilon^{2}}}\right)  \nonumber\\
&  -\frac{1}{\sqrt{1-\varepsilon^{2}}}\left[  b^{\dagger}b+\frac{\varepsilon
}{2}\left(  b^{\dagger2}+b^{2}\right)  \right]  .
\end{align}
For symmetric phase, $m=1$, $S_{\alpha}=\tilde{S}_{\alpha}$, we have%
\begin{align}
\left\langle S_{x}^{2}\right\rangle  &  =\left\langle \tilde{S}_{x}%
^{2}\right\rangle =\frac{N}{4}\sqrt{\frac{h-\gamma}{h-1}},\nonumber\\
\left\langle S_{y}^{2}\right\rangle  &  =\left\langle \tilde{S}_{y}%
^{2}\right\rangle =\frac{N}{4}\sqrt{\frac{h-1}{h-\gamma}},
\end{align}
while for broken phase, $m=h$, we need to rotate $\tilde{S}_{x}$ back to
$S_{x}$ as%
\begin{align}
S_{x} &  =\tilde{S}_{x}\cos\theta_{0}+\tilde{S}_{z}\sin\theta_{0}\nonumber\\
&  =h\tilde{S}_{x}+\sqrt{1-h^{2}}\tilde{S}_{x},
\end{align}
then we have
\begin{align}
\left\langle S_{x}^{2}\right\rangle = &  \sqrt{1-h^{2}}\left\langle \tilde
{S}_{z}^{2}\right\rangle +h\left\langle \tilde{S}_{x}^{2}\right\rangle
\nonumber\\
= &  \left(  \frac{N^{2}}{4}+\frac{N}{2}\right)  \left(  1-h^{2}\right)
\nonumber\\
&  +\frac{N}{4}\frac{\left(  1-\gamma\right)  h^{2}-\left(  2-h^{2}%
-\gamma\right)  \left(  1-h^{2}\right)  }{\sqrt{\left(  1-h^{2}\right)
\left(  1-\gamma\right)  }}.\label{Sx}%
\end{align}
We insert the above results into Eq. (\ref{sqz_fisher}). For polarized phase,
$h>1$, we have%
\begin{equation}
\xi_{1}^{2}=\chi^{2}=\sqrt{\frac{h-1}{h-\gamma}}<1.
\end{equation}
When $h$ is far from the critical point, $\xi_{1}^{2}$ and $\chi^{2}$ approach
to 1, then $\Delta\theta\sim\left(  \Delta\theta\right)  _{\text{SN}}$. For
broken phase, $h<1$, we get the spin squeezing,\emph{ }%
\begin{equation}
\xi_{1}^{2}=\sqrt{\frac{1-h^{2}}{1-\gamma}},
\end{equation}
while
\begin{equation}
\chi^{2}=\frac{N}{4\left\langle S_{x}^{2}\right\rangle }\simeq\frac{1}{\left(
N+2\right)  \left(  1-h^{2}\right)  }\simeq\frac{1}{N},
\end{equation}%
\begin{figure}
[ptb]
\begin{center}
\includegraphics[
height=3.5258in,
width=2.7754in
]%
{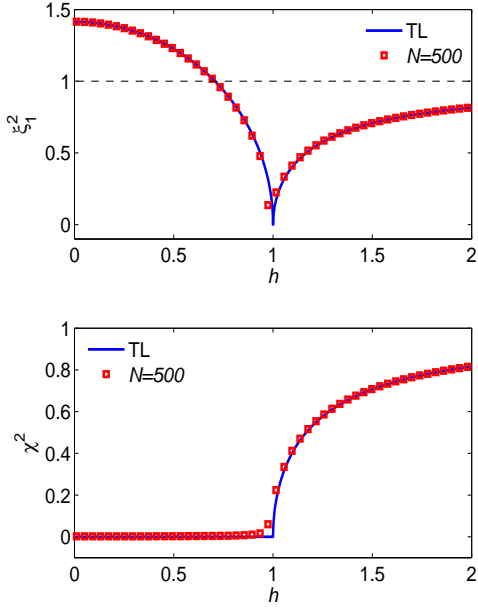}%
\caption{Comparison between analytical and numerical results for $\xi_{1}^{2}$
and $\chi^{2}$, in the case of $\gamma=1/2$. The solid line is analytical
results for the thermodynamic limit (TL).}%
\label{fig2}%
\end{center}
\end{figure}
Thus $\left(  \Delta\theta\right)  _{\text{QCR}}\simeq1/N$. When $h$
approaches to the critical point $h_{c}=1$, there are two limit processes in
Eq. (\ref{Sx}), that is $\left(  1-h\right)  $ tends to be zero and $N$ tends
to be infinity. To overcome this problem, we need to expand the Hamiltonian in
higher order of $1/N$. Fortunately, the finite size behaviors of the spin
squeezing and $\chi^{2}$ at the critical point can be derived by using the
results obtained in \cite{DusuelPRB71}, where the authors employ the continues
unitary transformations and get%
\begin{equation}
\frac{4\langle S_{x}^{2}\rangle}{N^{2}}\bigg{|}_{h=1}\sim\frac{a_{xx}^{\left(
0\right)  }}{N^{2/3}},~~\frac{4\langle S_{y}^{2}\rangle}{N^{2}}\bigg{|}_{h=1}%
\sim\frac{a_{yy}^{\left(  0\right)  }}{N^{4/3}},
\end{equation}
where $a_{xx}^{\left(  0\right)  }$ and $a_{yy}^{\left(  0\right)  }$ are
constant independent of $N$. Now we have%
\begin{equation}
\xi^{2}\big{|}_{h=1}\sim\frac{a_{yy}^{\left(  0\right)  }}{N^{2/3}},~~\chi
^{2}\big{|}_{h=1}\sim\frac{a_{xx}^{\left(  0\right)  }}{N^{2/3}},
\end{equation}
then for large $N$, $\xi^{2}$ and $\chi^{2}$ converge to zero as $1/N^{2/3}$,
and $\left(  \Delta\theta\right)  _{\text{QCR}}\sim1/N^{5/6}$.%
\begin{figure}
[ptb]
\begin{center}
\includegraphics[
height=2.1908in,
width=2.7195in
]%
{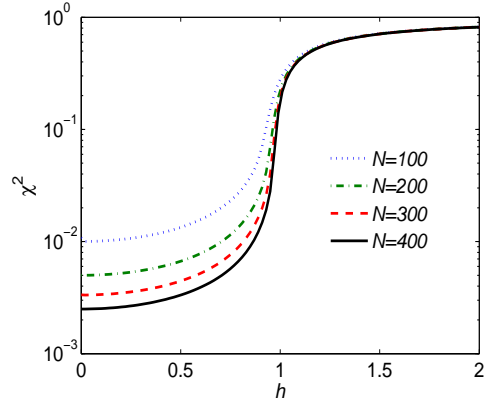}%
\caption{Plot of $\chi^{2}$ as a function of $h$ with $\gamma=1/2$
for different system sizes $N=100,200,300,400$, (from top to
bottom). $\chi^{2}$ is nearly independent of $N$ for $h>1$, while it
drops with the increasing of $N$
in the broken phase, $h<1$.}%
\label{fig4}%
\end{center}
\end{figure}
\begin{figure}
[ptb]
\begin{center}
\includegraphics[
height=2.1944in,
width=2.6796in
]%
{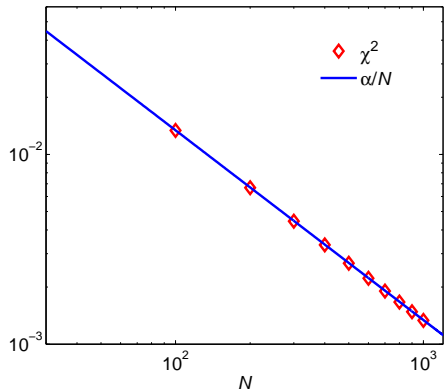}%
\caption{Scaling of $1/\chi^{2}$ as a function of $N$ with $h=\gamma=1/2$. The
diamonds are numerical results, while the solid line is a linear fit with
$\alpha$ a constant. It's obviously that $\chi^{2}\propto1/N$.}%
\label{fig3}%
\end{center}
\end{figure}

To verify these analytical prediction, in Figs. \ref{fig1} and \ref{fig2}, we
show numerical results for $\xi_{1}^{2}$ and $\chi^{2}$ as functions of $h$
with different $\gamma$ for finite size system. As shown in Fig.\ \ref{fig1},
in the symmetric phase, $\xi_{1}^{2}=\chi^{2}<1$, while in the broken phase,
$\chi^{2}$ and $\xi_{1}^{2}$ behaves very differently. In the most of
parameter range, $\chi^{2}<1$, which indicates entanglement, while for
$h\leq\sqrt{\gamma}$, $\xi_{1}^{2}\geq1$, and thus the system is not spin
squeezed (Fig. 1(d)). For the isotropic case, there is no spin squeezing. In
Fig. \ref{fig2}, we plot $\chi^{2}$ and $\xi_{1}^{2}$ for $N=500$ and the
thermaldynamical limit, and find the numerical results coincide well with the
analytical ones obtained by H-P transformation method.

As shown in Figs. \ref{fig4} and \ref{fig3}, $\chi^{2}$ is nearly independent
of larger $N$ in the symmetric phase, and approaches to $1/N$ as $h$ being
away from the critical point in the broken phase. Therefore, entanglement
characterized by $\chi^{2}$ is very different in the two phases, especially
when we treat them as resources for quantum estimation. In the symmetric
phase, as shown in Fig. \ref{fig4}, $\chi^{2}$ is nearly independent of $N$,
and the parameter sensitivity is at the level of $\left(  \Delta\theta\right)
_{\text{SN}}$, while in the broken phase, the ground states are more sensitive
in parameter. In Fig. (\ref{fig3}), we show numerical results for $\chi^{2}$
in the broken phase at $h=1/2$, $\gamma=1/2$, and we see clearly that
$\chi^{2}\propto1/N$. Therefore, the parameter estimation in the broken phase
is enhanced to the Heisenberg limit.

One can use the concurrence and entropy to quantity entanglement in the LMG
model and results are obtained in \cite{Barthel06,VidalPRA69,DusuelPRB71}. We
see that, both concurrence and entropy indicate well the presence of
entanglement, however, from them, we cannot tell whether the entanglement of
the ground state is useful in parameter estimation. From results of $\chi^{2}%
$, we can see that, the entanglements in these two phases are different
according to their performances in estimation. On one hand, we can use the
collective behavior of the LMG model to improve the phase estimation
precision, on the other hand, the differences of the parameter sensitivities
can be used to characterize and distinguish the two quantum phases.

\section{Conclusion}

We have analyzed $\chi^{2}$ and spin squeezing parameters $\xi_{i}^{2}$ in the
ground state of the LMG model. For isotropic case, the Hamiltonian is diagonal
in Dicke states, for which we have $\xi_{2}^{2}\geq\xi_{1}^{2}=1/\chi^{2}$.
For anisotropic case, our results indicate that, $\chi^{2}$ classifies states
in different phases in the sense of quantum phase estimation. Hence we can use
$\chi^{2}$ to distinguish and characterize the behaviors of the two phases of
the LMG model. In the symmetric phase, $\chi^{2}$ is independent of $N$ and
approaches to 1 with the increasing of $h$, thus $\left(  \Delta\theta\right)
_{\text{QCR}}\sim\left(  \Delta\theta\right)  _{\text{SN}}$, that is just a
bit lower than the shot-noise limit. In the broken phase, we find $\chi
^{2}\simeq1/N$ and $\left(  \Delta\theta\right)  _{\text{QCR}}\simeq1/N$,
which attains the Heisenberg limit.

Fisher information, being related to the Cramer-Rao inequality, is
used to measure how much information that we know about some certain
parameters in a probability distribution. From present results, we
see that Fisher information can also characterize the QPT, by
distinguishing the entangled ground states in the sense of parameter
sensitivity. This approach is promising and expected to be
applicable to other spin systems undergoing a QPT.

\begin{acknowledgments}The authors thanks helpful discussions with
S. J. Gu, C. P. Sun, W. F. Liu, and H. N. Xiong. This work is
supported by NSFC with grant No.10874151, NFRPC with grant No.
2006CB921205; Program for New Century Excellent Talents in
University (NCET).
\end{acknowledgments}

\end{document}